\documentclass[preprint,showpacs,preprintnumbers,amsmath,amssymb,nofootinbib]{revtex4}
\usepackage{graphicx}
\usepackage{dcolumn}% Align table columns on decimal point
\usepackage{bm}% bold math
 
\begin{document}
\title{\bf Exterior Differential Systems for Field Theories}
\author{Frank B. Estabrook}
\email{frank.b.estabrook@jpl.nasa.gov}
\affiliation{Jet Propulsion Laboratory, California Institute of Technology, Pasadena, CA 91109}
                                
\date{\today}

\begin{abstract}Exterior Differential Systems (EDS) and Cartan forms, set in the state space of field variables taken together with four space-time variables, are formulated for classical gauge theories of Maxwell and SU(2) Yang-Mills fields minimally coupled to Dirac spinor multiplets. Cartan character tables are calculated, showing whether the EDS, and so the Euler-Lagrange partial differential equations, is well-posed.  The first theory, with 22 dimensional state space (10 Maxwell field and potential components and  8  components of a Dirac field), anticipates QED.  In the second, non-Abelian, case (30 Yang-Mills field components and 16 Dirac), only if three additional "ghost" fields are included (15 more scalar variables) is a well-posed EDS found. This classical formulation anticipates the need for introduction of Fadeev-Popov ghost fields in the quantum standard model.
\end{abstract}  
\maketitle
\baselineskip= 14pt

\section{Introduction.  Gauge fields} 

Well posed exterior differential systems (EDS) for Maxwell's vacuum
(source free) equations, and for their gauge or Yang-Mills generalizations,
have previously been given \cite{esta}.  In the first case the forms
generating the system live in a 14 dimensional "state space": ten
variables spanning potential and field components, together with four
independent space-time variables.  For SU(2) Yang-Mills the state
space dimension rises to thirty-four.  Cartan-K\"{a}hler theory characterizes solutions of an EDS as those
(in the present case) four dimensional submanifolds of state space
that are the maximal null set, or solution, of a closed ideal of
exterior differential forms, the EDS \cite{Ivey}. These are general solutions of
sets of first order partial differential equations.

In the Maxwell case the ideal is generated by a single 2-form $\theta$ and its
exterior derivative $d\theta$, together with a an additional "dynamic"
3-form $\psi$.  $\theta$ and $\psi$ are functions of the state space
coordinates, $F_{ij}, A_i, x,y,z,t$:
\begin{eqnarray}
\theta & = & \text{dA}_1\wedge\text{dx}+\text{dA}_2\wedge\text{dy}+\text{dA}_3\wedge\text{dz}+\text{dA}_4\wedge\text{dt}-F_{14} \text{dx}\wedge\text{dt} \nonumber \\
       &   & \mbox{}-F_{12} \text{dx}\wedge\text{dy}+F_{31} \text{dx}\wedge\text{dz}-F_{24} \text{dy}\wedge\text{dt} 
-F_{23} \text{dy}\wedge\text{dz}-F_{34} \text{dz}\wedge\text{dt}
\end{eqnarray}
\begin{eqnarray}
\psi & = & - \text{dF}_{12}\wedge\text{dz}\wedge\text{dt}+\text{dF}_{14}\wedge\text{dy}\wedge\text{dz}-\text{dF}_{23}\wedge\text{dx}\wedge\text{dt}\nonumber \\&   &-\text{dF}_{24}\wedge\text{dx}\wedge\text{dz}  -\text{dF}_{31}\wedge\text{dy}\wedge\text{dt}+\text{dF}_{34}\wedge\text{dx}\wedge\text{dy}
\end{eqnarray}

In the Yang-Mills SU(2) generalization of this we have three such pairs of forms, $\theta^a$ and $\psi_a$, $a=1..,.3$,  Their definitions in terms of potentials $A^a_i$ and fields $F^a_{ij}$ are given in Ref. 1.  This 2-form/3-form structure of an EDS characterizes so-called gauge theories.  On the other hand, EDS's generated only by "contact" 1-forms and 4-forms are called multicontact systems \cite{Bry} \cite{Gim}.  These include the scalar field theory we will discuss in Section II and source free Dirac theory discussed in Section III (although there no contact forms are used).  In subsequent sections we discuss EDS's for minimally coupled field theories where generating forms of all ranks occur.

The set of generators of an EDS--1-forms, 2-forms, 3-forms, and so on-- must
be closed under exterior differentiation; Cartan's non-perturbative theory of the existence and structure of
solutions of an EDS requires calculation of a series of so-called Cartan
characters, integers determined at a single, generic, point of N-dimensional state
space from the ranks of a series of nested linear sets of equations
sequentially set using auxiliary vectors.  The characters are denoted
$s_i$, $i=0,1,2...$ ; the theory shows how these calculations must stop
at, say, n-1, giving n as the dimension of the solution submanifold.  If the signature is right, the sum of these characters is the number of evolution equations in the equivalent set of partial differential equations.  The characters also categorize the constraints, or integrability conditions, encountered in numerical integration.  A final Cartan character is conventionally computed, denoted $s_n$, it is the number of arbitrary functions that enter the general solution;
if this is non-zero, in field theory it is customarily called the degree of gauge
freedom.

An EDS belongs to a variational principle if there is a Cartan n-form equivalent to a
Lagrangian density \cite{Got}; its exterior derivative, the so-called
multisymplectic n+1-form, contracted with
\emph{arbitrary} vectors must give n-forms in the ideal of the generators of the EDS (this is arbitrary \emph{variation} of the Cartan 4-form).  If, using only these, the character $s_{n-1}$ is less than $N-n$ this signals the need for lower rank forms to generate a well posed EDS.  If it  can thus be completed, the EDS then codes the Euler-Lagrange partial differential equations of the variational principle, together with all their integrability conditions or constraints. \cite{cendra} 

The multisymplectic 5-form
for Maxwell is the exterior product $\theta \wedge \psi$, and that for
the other free field Yang-Mills gauge theories $\theta^a\wedge\psi_a$, and these factors generate the EDSs.  The multisymplectic form for multicontact EDSs can be formed from just exterior products of 1-forms and 4-forms. As we will see in Section III, the Cartan form for the Dirac equation has a special relation to the EDS, and the multisymplectic form does not factor but $s_{n-1}$ is just $N-n$; no potential fields are introduced, and the EDS is generated solely by 4-forms.  There even exist systems, such as the Hilbert Lagrangian for vacuum general relativity imbedded in flat 10-space, where the multisymplectic form factors in either way, giving different well-posed EDSs that may describe the possibility of gravitational field phase change \cite{grav} \cite{Paston}.

In calculating the various new sets of Cartan characters reported here we used a small suite of Mathematica
programs written by the late H. D. Wahlquist, called AVF (Algebra
Valued, or indexed, Forms).  They have been carefully edited by Jos\'{e} M Mart\'{\i}n-Garcia and are now available on the xAct website
\cite{xAct}.  We will report the characters obtained for an EDS in an array $N(s_0,s_1...s_{n-1})n+s_n$.
                  
An important test that is checked sequentially during the
calculation of the characters of an EDS is that those n (here four)
state space variables one anticipates taking as independent when writing an
equivalent set of partial differential equations remain so in the
solution submanifold and can indeed be used.  We have refered to this
property as being ''well-posed".  Cartan calls such variables as being
"in involution" or "involutory".  A well-posed EDS then satisfies "Cartan's test" \cite{Ivey}.  In more recent language, we want solutions to be cross sections of a a bundle over an n-dimensional base.  The AVF suite checks well-posedness beginning with the
first n coordinates of state space that are entered in its coordinate
list, so we always enter $x,y,z,t$ first in an AVF calculation.

In Ref. [1] the tables of Cartan characters, $N[s_0,s_1,s_2,s_3]4 + s_4$ of vacuum Maxwell and SU(2) Yang-Mills theories in four dimensions were reported as respectively $14[0,1,3,5]4+1$, and $34[0,3,9,15]4+3$.  The coordinates  $x,y,z,t$ are in involution, and the degrees of gauge freedom one and three, as shown. An EDS for three coupled scalar fields is given in Section II, 19(3,3,3,6)4, and a single Dirac spinor field in Section III has 12[0,0,0,8]4.   

In Sections IV and V we treat field theories with Cartan and multisymplectic forms that "minimally" couple suitable multiplets of the free Dirac equation to the EDS's for Maxwell and SU(2)
Yang-Mills.  We calculate their Cartan character tables. and check involutivity.  Only
the first of these, QED, immediately proves to be well-posed; it also has one degree of gauge freedom.  For the SU(2)  field theory we find well-posedness only if the two sets of Dirac fields are further supplemented by three scalar "ghost" fields. The "ghosts"  enter the Cartan form (Lagrangian density) only as coupled to the Yang-Mills or gauge fields, 
\section{Scalar Fields}
We treat a multicontact EDS for a multiplet of three. The 19 dimensional state space is spanned by 15 fields $\rho _1,\rho _2,\rho _3,\rho _{11},\rho _{12},\rho _{13},\rho _{14},\rho _{21},\rho _{22},\rho _{23},\rho _{24},\rho _{31},\rho _{32},\rho _{33},\rho _{34}$ together with $x,y,z,t$.  The multicontact EDS is generated by three contact 1-forms and their exterior derivatives
\begin{align}
\sigma_1 & = \text{d$\rho $}_1-\rho _{11}\text{dx}_1-\rho _{12}\text{dx}_2-\rho _{13}\text{dx}_3-\rho _{14}\text{dx}_4\\
\sigma_2 & = \text{d$\rho $}_2-\rho _{21}\text{dx}_1-\rho _{22}\text{dx}_2-\rho _{23}\text{dx}_3-\rho _{24}\text{dx}_4\\
\sigma_3 & = \text{d$\rho $}_3-\rho _{31}\text{dx}_1-\rho _{32}\text{dx}_2-\rho _{33}\text{dx}_3-\rho _{34}\text{dx}_4
\end{align}
and three 4-forms to carry the dynamic content:
\begin{eqnarray} 
\Sigma_1 & = & \lambda ^2 \rho _1 \left(-\mu ^2+\rho _1^2+\rho _2^2+\rho _3^2\right) \text{dx}_1\wedge \text{dx}_2\wedge \text{dx}_3\wedge \text{dx}_4+\text{dx}_1\wedge \text{dx}_2\wedge \text{dx}_3\wedge \text{d$\rho $}_{14}\nonumber \\  &   & 
  -\text{dx}_1\wedge \text{dx}_2\wedge \text{dx}_4\wedge \text{d$\rho $}_{13}+\text{dx}_1\wedge \text{dx}_3\wedge \text{dx}_4\wedge \text{d$\rho $}_{12}-\text{dx}_2\wedge \text{dx}_3\wedge \text{dx}_4\wedge \text{d$\rho $}_{11}
\end{eqnarray}
\begin{eqnarray}
\Sigma_2 & = &\lambda ^2 \rho _2 \left(-\mu ^2+\rho _1^2+\rho _2^2+\rho _3^2\right) \text{dx}_1\wedge \text{dx}_2\wedge \text{dx}_3\wedge \text{dx}_4+\text{dx}_1\wedge \text{dx}_2\wedge \text{dx}_3\wedge \text{d$\rho $}_{24}
\nonumber \\  &   & 
 -\text{dx}_1\wedge \text{dx}_2\wedge \text{dx}_4\wedge \text{d$\rho $}_{23}+\text{dx}_1\wedge \text{dx}_3\wedge \text{dx}_4\wedge \text{d$\rho $}_{22}-\text{dx}_2\wedge \text{dx}_3\wedge \text{dx}_4\wedge \text{d$\rho $}_{21}
\end{eqnarray}
\begin{eqnarray}
\Sigma_3 & = &\lambda ^2 \rho _3 \left(-\mu ^2+\rho _1^2+\rho _2^2+\rho _3^2\right) \text{dx}_1\wedge \text{dx}_2\wedge \text{dx}_3\wedge \text{dx}_4+\text{dx}_1\wedge \text{dx}_2\wedge \text{dx}_3\wedge \text{d$\rho $}_{34}
\nonumber \\  &   &  -\text{dx}_1\wedge \text{dx}_2\wedge \text{dx}_4\wedge \text{d$\rho $}_{33}+\text{dx}_1\wedge \text{dx}_3\wedge \text{dx}_4\wedge \text{d$\rho $}_{32}-\text{dx}_2\wedge \text{dx}_3\wedge \text{dx}_4\wedge \text{d$\rho $}_{31}
\end{eqnarray}

AVF calculates the Cartan integer table to be $19(3,3,3,6)4$ with no gauge freedom. $x,y,z,t$ are in involution, so equivalent partial differential
equations that adopt them as independent variables are well-posed..  The exact multisymplectic form is
\begin{eqnarray}
\text{d$\Lambda $} = \sigma _1\wedge\Sigma _1+\sigma _2\wedge\Sigma _2+\sigma _3\wedge\Sigma _3
\end{eqnarray}
Integrating this by parts gives a Cartan form, unique up to an exact 4-form:

\begin{eqnarray}
\Lambda & = & -\rho _1 \left(\text{d$\rho $}_{11} \wedge \eta _1+\text{d$\rho $}_{12} \wedge \eta _2+\text{d$\rho $}_{13} \wedge \eta _3
+\text{d$\rho $}_{14} \wedge \eta _4\right) \nonumber \\ &    &
-\rho _2 \left(\text{d$\rho $}_{21} \wedge \eta _1+\text{d$\rho $}_{22} \wedge \eta _2+\text{d$\rho $}_{23} \wedge \eta _3+\text{d$\rho $}_{24} \wedge \eta _4\right)\nonumber \\ &   &
-\rho _3 \left(\text{d$\rho $}_{31} \wedge \eta _1+\text{d$\rho $}_{32} \wedge \eta _2
+\text{d$\rho $}_{33} \wedge \eta _3+\text{d$\rho $}_{34} \wedge \eta _4\right) \nonumber \\ &   &
-(1/2)\Bigl(\rho _{11}^2+\rho _{12}^2+\rho _{13}^2+\rho _{14}^2
+\rho _{21}^2+\rho _{22}^2+\rho _{23}^2+\rho _{24}^2
+\rho _{31}^2 + \rho _{32}^2+\rho _{33}^2+\rho _{34}^2 \nonumber \\ &   &
+\lambda^2  \left(\rho _1^2+\rho _2 ^2+\rho _3^2-\mu^2 \right)^2 /2\Bigr)\eta
\end{eqnarray}

where a useful notation for the basic 3-forms and 4-form we will use henceforth is
\begin{eqnarray}
\eta _0 & = & dx\wedge dy \wedge dz \\
\eta _1 & = & dt \wedge dz\wedge dy \\
\eta _2 & = & dz \wedge dt \wedge dx \\
\eta _3 & = & dx \wedge dt \wedge dy \\
\eta  & = & dt \wedge dx \wedge \ dy \wedge dz
\end{eqnarray}

As we have emphasized $\Lambda$ lives in the 19 dimensional state space.  Field theorists however customarily anticipate the bundle structure, substitute back into into such Cartan forms the contact 1-forms from the EDS, $\sigma _1,\sigma _2,\sigma _3$, eliminating $d\rho _1,d\rho _2,d\rho _3$, and writing the $\rho_{ij}$ as pullbacka $\partial_j \rho_i=\rho_{i,j}$.  So treated, the Cartan 4-form becomes an expression $ L \eta$ and the Lagrangian density functional $L$, or $(-T+V)$, is used in variational calculus.  In this case we have
\begin{eqnarray}
L/2 & = & - \rho _{1,1}^2- \rho _{1,2}^2- \rho _{1,3}^2- \rho _{1,4}^2- \rho _{2,1}^2- \rho _{2,2}^2- \rho _{2,3}^2- \rho _{2,4}^2- \rho _{3,1}^2 \nonumber \\
&   & \mbox{}- \rho _{3,2}^2- \rho _{3,3}^2- \rho _{3,4}^2+ \lambda ^2\left.\left(\rho _1^2+\rho _2^2+\rho _3^2- \mu ^2\right)^2\right/2
\end{eqnarray} 
All the other Cartan forms given below can be similarly treated to become Lagrangian densities.

Other important associated structures to an EDS in state space are conservation
laws (or currents).  These are 3-forms, \emph{not} in the ideal,  whose exterior derivatives \emph{are}
in the ideal, the EDS. Here there are three of these:

\begin{equation}
J_{1s}=\left(\rho _{21} \rho _3 \eta _1+\rho _{22} \rho _3 \eta _2+\rho _{23} \rho _3 \eta _3+\rho _{24} \rho _3 \eta _0\right)-\left(\rho _{31} \rho _2 \eta _1+\rho _{32} \rho _2 \eta _2+\rho _{33} \rho _2 \eta _3+\rho _{34} \rho _2 \eta _0\right)
\end{equation}

\begin{equation}
J_{2s}=\left(\rho _{31} \rho _1 \eta _1+\rho _{32} \rho _1 \eta _2+\rho _{33} \rho _1 \eta _3+\rho _{34} \rho _1 \eta _0\right)-\left(\rho _{11} \rho _3 \eta _1+\rho _{12} \rho _3 \eta _2+\rho _{13} \rho _3 \eta _3+\rho _{14} \rho _3 \eta _0\right)
\end{equation}

\begin{equation}
J_{3s}=\left(\rho _{11} \rho _2 \eta _1+\rho _{12} \rho _2 \eta _2+\rho _{13} \rho _2 \eta _3+\rho _{14} \rho _2 \eta _0\right)-\left(\rho _{21} \rho _1 \eta _1+\rho _{22} \rho _1 \eta _2+\rho _{23} \rho _1 \eta _3+\rho _{24} \rho _1 \eta _0\right)
\end{equation}

\section{Dirac Fields} 

The eight Dirac Equations (sic) are for eight fields that are
functions of four independent variables.  They are usually written in
compressed spinor notation, but we need explicit state space
coordinates, say $X_i + iY_i$ for each complex Dirac spinor
component, and together with spacetime coordinates $x,y,z,t$ they span
a twelve dimensional state space.  As an EDS it has notably different structure from that for Maxwell
and the gauge theories; it is a sub case of a multicontact system,
coded as eight 4-forms, say $\Xi_i$ and $\Psi_i$, and the character
table is just $12(0,0,0,8)4$ with no constraints and no gauge freedom.
Unlike the Maxwell and Yang-Mills systems no potential fields were needed to
be adjoined to achieve a variational principle. Also,
its generalization appears to be just the use of multiple copies, or
multiplets.
\begin{align}
\Xi _1 & = \text{dX}_1 \wedge \eta _0 + \text{dX}_0 \wedge \eta_1 
+ \text{dY}_0 \wedge \eta _2 + \text{dX}_3 \wedge \eta_3 - m Y_1 \eta 
\\
\Psi _1 & = \text{dY}_1\wedge \eta _0+\text{dY}_0\wedge \eta
_1-\text{dX}_0\wedge \eta _2+\text{dY}_3\wedge \eta_3+m X_1\eta 
\\
\Xi _2 & = \text{dX}_2\wedge \eta _0+\text{dX}_3\wedge \eta
_1-\text{dY}_3\wedge \eta _2-\text{dX}_0\wedge \eta_3-m Y_2\eta 
\\
\Psi _2 & = \text{dY}_2\wedge \eta _0+\text{dY}_3\wedge \eta
_1+\text{dX}_3\wedge \eta _2-\text{dY}_0\wedge \eta_3+m X_2\eta 
\\
\Xi _3 & = -\text{dX}_3\wedge \eta _0-\text{dX}_2\wedge \eta
_1-\text{dY}_2\wedge \eta _2-\text{dX}_1\wedge \eta_3-m Y_3\eta 
\\
\Psi _3 & = -\text{dY}_3\wedge \eta _0-\text{dY}_2\wedge \eta
_1+\text{dX}_2\wedge \eta _2-\text{dY}_1\wedge \eta_3+m X_3\eta 
\\
\Xi _0 & = -\text{dX}_0\wedge \eta _0-\text{dX}_1\wedge \eta
_1+\text{dY}_1\wedge \eta _2+\text{dX}_2\wedge \eta_3-m Y_0\eta 
\\
\Psi _0 & = -\text{dY}_0\wedge \eta _0-\text{dY}_1\wedge \eta
_1-\text{dX}_1\wedge \eta _2+\text{dY}_2\wedge \eta_3+m X_0\eta
\end{align}
The $x, y, z, t$ are in involution so equivalent partial differential
equations using them as independent variables--the Dirac equation--are well posed.

A Lagrangian density, hence a Cartan 4-form, say $\Lambda_D$, for the Dirac
partial differential set is well known;  in our notation it is \begin{equation}
\Lambda_D=(X_1\Psi _1-Y_1\Xi _1+X_2\Psi _2-Y_2\Xi _2+X_3\Psi _3-Y_3\Xi _3+X_0\Psi _0-Y_0\Xi _0)/2\end{equation}
  $\Lambda_D$ is of course not unique, but
its exterior derivative $d\Lambda_D$, the
multisymplectic 5-form, is.  Expanding this, for the Dirac system we have
\begin{eqnarray}
\text{d$\Lambda_D $} & = & 
\left(\text{dX}_1 \wedge \text{dY}_1 + \text{dX}_2 \wedge
  \text{dY}_2 +\text{dX}_3 \wedge \text{dY}_3 + \text{dX}_0 \wedge
  \text{dY}_0 \right)
\wedge \eta_0 
\nonumber
\\
& + & \left(\text{dX}_1 \wedge \text{dY}_0 + \text{dX}_0 \wedge
  \text{dY}_1 
+ \text{dX}_3 \wedge \text{dY}_2 +\text{dX}_2 \wedge \text{dY}_3 \right)
\wedge  \eta_1 
\nonumber
\\
& + & \left( \text{dX}_0 \wedge \text{dX}_1 +
  \text{dY}_0 \wedge
  \text{dY}_1 + \text{dX}_2 \wedge \text{dX}_3 + \text{dY}_2 \wedge
  \text{dY}_3 \right)
\wedge \eta_2 
\nonumber
\\
& + & \left(\text{dX}_1 \wedge \text{dY}_3 + \text{dX}_3 \wedge
  \text{dY}_1 -\text{dX}_2 \wedge \text{dY}_0 - \text{dX}_0 \wedge
  \text{dY}_2 \right)
\wedge \eta_3 
\nonumber
\\
& + &  m \left(X_1 \text{dX}_1 +Y_1 \text{dY}_1 + X_2
  \text{dX}_2 + Y_2 \text{dY}_2 - X_3 \text{dX}_3 - Y_3 \text{dY}_3 
\right.
\nonumber
\\
& & \left. - X_0   \text{dX}_0-Y_0 \text{dY}_0 \right) \wedge \eta
\end{eqnarray}
This can also be written as
\begin{eqnarray}
2 \text{d$\Lambda_D $} & = & \text{dX}_1\wedge\Psi _1-\text{dY}_1\wedge\Xi _1+\text{dX}_2\wedge\Psi _2-\text{dY}_2\wedge\Xi _2\nonumber \\
&   & +\text{dX}_3\wedge\Psi _3-\text{dY}_3\wedge\Xi _3+\text{dX}_0\wedge\Psi _0-\text{dY}_0\wedge\Xi _0 \nonumber \\
&   & + m \left.d\left[X_1^2+Y_1^2+X_2^2+Y_2^2-X_3^2-Y_3^2-X_0^2-Y_0^2\right]\wedge\eta /2\right.
\end{eqnarray}
An arbitrary vector in state space, say X, contracted on this 5-form yields a
4-form that is in the EDS generated by the $\Xi_i$ and $\Psi_i$.
  The Dirac system, spanned only by 4-forms, is also seen to be further anomalous in that its Cartan form (28) and functional Lagrangian vanish when evaluated on solutions of the EDS they generate variationally.

For the Dirac ideal the conserved current $J$ is
\begin{eqnarray}
J &=& \left(X_1^2 + Y_1^2 +X_2^2 + Y_2^2
+X_3^2 + Y_3^2 + X_0^2 + Y_0^ 2\right)
\eta_0 + 2 \left(X_1 X_0 + Y_1 Y_0 + X_2 X_3+Y_2 Y_3\right)\eta_1
\nonumber
\\
& + & 2 \left(-Y_1 X_0+X_1 Y_0-X_2 Y_3+Y_2 X_3\right)
\eta_2 + 2 \left(X_1 X_3+Y_1 Y_3-X_2 X_0-Y_2 Y_0\right)\eta_3
\end{eqnarray}
from which
\begin{equation}
dJ=2\left( X_1\Xi _1+ Y_1\Psi _1+\text{  }X_2\Xi _2+ Y_2\Psi _2+\text{  }X_3\Xi _3+ Y_3\Psi _3+\text{  }X_0\Xi _0+ Y_0\Psi _0\right)
\end{equation}
\section{Maxwell-Dirac Theory}

In the classical version of QED, the outer product of the potential
1-form $A=A_i dx^i$ of Maxwell theory and of the conserved current 3-form $J$ of Dirac
theory is taken as a "minimal" coupling term added to the sum of the
two respective Cartan 4-forms.  An EDS formulating this is implied in an elegant and
overlooked paper of Barut, Moore and Piron\cite{Bar}.  They first neatly expound how the
canonical Cartan 1-form and its exterior derivative 2-form, or symplectic structure, work in
classical mechanics and their generalization to the higher dimension
of field theory.  They discuss the straightforward use of Cartan forms in state space vs. the subtlety of   variational/functional calculus.  They
present the Cartan $\Lambda$ form in the 22 dimensional state space of coupled Maxwell-Dirac
theory. and propose an EDS from contractions of an arbitrary vector on the multisymplectic 5-form $d\Lambda$.  It is generated by 18 4-forms and we calculate its character table to be $22(0,0,0,15,0,0)7$, and moreover not in involution.  Barut et al state that it is a simple matter then to write the field theoretic partial differential equations, and indeed it is apparent from inspection  that the lower rank gauge 2-form $\theta$ and 3-forms $d\theta$ and $\psi + J$ , like those of the pure Maxwell EDS, can be added to just 8 4-forms as generators.  We calculate this specialized EDS to have characters $22(0,1,3,13)4+1$ with $x, y, z, t$ now in involution, and one degree of gauge freedom.  It of course directly yields the partial differential equations Barut et al discuss.
       
The
motivation of Barut et al was to go on to show how the Schr\"{o}dinger and Maxwell-Schr\"{o}dinger equations emerge as direct limits of the relativistic field equations.

Taking note of the identity $J{\wedge}(\theta-\text{d}A)=0$ the multisymplectic 5-form for the system can be written to show how the
coupling has affected both the gauge and spinor EDSs combined in this field theory; the 3-form second
factor in the Maxwell multisymplectic 5-form $-\theta{\wedge}\psi$ now
includes a $J$ current form, while the 5-form of the spinor field now
includes $A{\wedge}dJ$ (in the 4-forms of the EDS and the resulting partial
differential equations it is this last modification that appears as a type of
"covariant" differentiation):
\begin{equation}
\text{d$\Lambda $}= -\theta \wedge ( \psi + J)+\text{$d\Lambda_D $}+A\wedge dJ
\end{equation}
\begin{equation}
\text{EDS}:               \theta,  d\theta,  ( \psi + J),  d( \psi + J), \text{$X \bullet d\Lambda $}
\end{equation}

\begin{equation}
\text{$22(0,1,3,13)4+1$}
\end{equation}
X is an arbitrary vector field in the state space.

We learn from this example that finding a well-posed EDS from a given field theoretic Lagrangian or Cartan form is not quite straightforward, that the presence of lower rank generating forms is signaled by the algebraic structure of the Cartan form, and by how the to-be-independent variables enter.  cf. Ref.[6].  The character table of a possible EDS must then always be calculated and well-posedness confirmed.  In the following simple field theory, involving textbook Yang-Mills generalizations of Maxwell-Dirac theory, and using Equation (33) as a guide, we report success of this program only if additional currents from "ghost" fields are introduced.

\section{Coupled SU(2) Gauge and Dirac Fields}

It is surprising that in addition to 3-forms like Eq. (30) multiple copies of the Dirac equations allow
joint nontrivial conservation laws, though this is well known to particle theorists.  For example if we have two sets
of variables, say $X_{i,1},Y_{i,1}$ and $X_{i,2},Y_{i,2}$ and an EDS
generated by two \emph{independent} sets of Dirac 4-forms, in a 20
dimensional state space,  then there are three non
trivial joint conserved 3-forms, $J_a$, which we next write.  It can
be verified that the forms $dJ_a$ are in the EDS generated by two independent
copies of Equations (20)-(27), 20(0,0,0,16)4.  We use a recipe \cite{Fad} based on
representation theory of U(2) (the Pauli matrices):

\begin{eqnarray}
J_1 & = & -\left(X_{1,1}X_{1,2}+Y_{1,1}Y_{1,2}+X_{2,1}X_{2,2}+Y_{2,1}Y_{2,2}+X_{3,1}X_{3,2}+Y_{3,1}Y_{3,2}+X_{0,1}X_{0,2}+Y_{0,1}Y_{0,2}\right)\eta_0
\nonumber
\\
& - & \left(X_{1,1} X_{0,2}+Y_{1,1} Y_{0,2}+X_{2,1} X_{3,2}+Y_{2,1}
  Y_{3,2} + X_{1,2} X_{0,1}+Y_{1,2} Y_{0,1}+X_{2,2} X_{3,1}+Y_{2,2}
  Y_{3,1}\right)\eta_1
\nonumber
\\
& - & \left(-Y_{1,1} X_{0,2}+X_{1,1} Y_{0,2}-X_{2,1} Y_{3,2}+Y_{2,1}
  X_{3,2}-Y_{1,2} X_{0,1}+X_{1,2} Y_{0,1}-X_{2,2} Y_{3,1}+Y_{2,2}
  X_{3,1}\right)\eta_2
\nonumber
\\
& - & \left(X_{1,1} X_{3,2}+Y_{1,1} Y_{3,2}-X_{2,1} X_{0,2}-Y_{2,1}
  Y_{0,2}+X_{1,2} X_{3,1}+Y_{1,2} Y_{3,1}-X_{2,2} X_{0,1} \right.
\nonumber
\\
&& \left. - Y_{2,2} Y_{0,1}\right)\eta_3
\end{eqnarray}

\begin{eqnarray}
J_2 & = &
\left(-Y_{1,1}X_{1,2}+X_{1,1}Y_{1,2}-Y_{2,1}X_{2,2}+X_{2,1}Y_{2,2}-Y_{3,1}X_{3,2}+X_{3,1}Y_{3,2}-Y_{0,1}X_{0,2}+X_{0,1}Y_{0,2}\right)\eta_0
\nonumber
\\
& + &
\left(-Y_{1,1}X_{0,2}+X_{1,1}Y_{0,2}+X_{2,1}Y_{3,2}-Y_{2,1}X_{3,2}+X_{3,1}Y_{2,2}-Y_{3,1}X_{2,2}+X_{0,1}Y_{1,2}-Y_{0,1}X_{1,2}\right)\eta_1
\nonumber
\\
& + &
\left(-X_{1,1}X_{0,2}-Y_{1,1}Y_{0,2}+X_{2,1}X_{3,2}+Y_{2,1}Y_{3,2}-X_{3,1}X_{2,2}-Y_{3,1}Y_{2,2}+X_{0,1}X_{1,2}+Y_{0,1}Y_{1,2}\right)\eta_2
\nonumber
\\
& + &
\left(X_{3,1}Y_{1,2}-Y_{3,1}X_{1,2}-X_{0,1}Y_{2,2}+Y_{0,1}X_{2,2}+X_{1,1}Y_{3,2}-Y_{1,1}X_{3,2}-X_{2,1}Y_{0,2}
  \right.
\nonumber
\\
&& \left. + Y_{2,1}X_{0,2}\right)\eta_3
\end{eqnarray}

\begin{eqnarray}
J_3 & = & -\left(X_{1,1}^2 + Y_{1,1}^2 + X_{2,1}^2 + Y_{2,1}^2
+ X_{3,1}^2 + Y_{3,1}^2 + X_{0,1}^2 + Y_{0,1}^2 - X_{1,2}^2 - Y_{1,2}^2
- X_{2,2}^2 - Y_{2,2}^2 \right.
\nonumber
\\
&& \left. - X_{3,2}^2 - Y_{3,2}^2 - X_{0,2}^2 -
Y_{0,2}^2 \right)
\frac{\eta_0}{2} - \left(X_{1,1} X_{0,1} + Y_{1,1} Y_{0,1} + X_{2,1}
  X_{3,1} + Y_{2,1} Y_{3,1} - X_{1,2} X_{0,2} \right.
\nonumber
\\
&& \left. - Y_{1,2} Y_{0,2} -  X_{2,2} X_{3,2} - Y_{2,2} Y_{3,2}\right)\eta_1
- \left(-Y_{1,1} X_{0,1} + X_{1,1} Y_{0,1} - X_{2,1} Y_{3,1} + Y_{2,1}
  X_{3,1}  \right.
\nonumber
\\
&& \left. + Y_{1,2} X_{0,2} -X_{1,2} Y_{0,2}+X_{2,2} Y_{3,2}-Y_{2,2}
  X_{3,2}\right)\eta_2
- \left(X_{1,1} X_{3,1} + Y_{1,1} Y_{3,1} - X_{2,1} X_{0,1}  \right.
\nonumber
\\
&& \left.- Y_{2,1}
  Y_{0,1} - X_{1,2} X_{3,2}-Y_{1,2} Y_{3,2}+X_{2,2} X_{0,2}
+ Y_{2,2} Y_{0,2}\right)\eta_3  
\end{eqnarray}

More succinctly we calculate 
\begin{eqnarray}
 d[{J_1}]+X_{1,1}\Xi_{1,2}+X_{1,2}\Xi_{1,1}+Y_{1,1}\Psi_{1,2}+Y_{1,2}\Psi_{1,1}+X_{2,1}\Xi_{2,2}\nonumber \\+X_{2,2}\Xi_{2,1}+Y_{2,1}\Psi_{2,2} +Y_{2,2}\Psi_{2,1}+X_{3,1}\Xi_{3,2}+X_{3,2}\Xi_{3,1}+\nonumber \\Y_{3,1}\Psi_{3,2}+Y_{3,2}\Psi_{3,1}+X_{0,1}\Xi_{4,2}+X_{0,2}\Xi_{4,1}+Y_{0,1}\Psi_{4,2}+Y_{0,2}\Psi_{4,1}=0
\end{eqnarray}
\begin{eqnarray}
 d{J_2}] +Y_{1,1}\Xi_{1,2}-Y_{1,2}\Xi_{1,1}-X_{1,1}\Psi_{1,2}+X_{1,2}\Psi_{1,1}+ Y_{2,1}\Xi_{2,2}\nonumber \\-Y_{2,2}\Xi_{2,1}-X_{2,1}\Psi_{2,2}+X_{2,2}\Psi_{2,1}+ Y_{3,1}\Xi_{3,2}-Y_{3,2}\Xi_{3,1}-\nonumber \\X_{3,1}\Psi_{3,2}+X_{3,2}\Psi_{3,1}+ Y_{0,1}\Xi_{4,2}-Y_{0,2}\Xi_{4,1}-X_{0,1}\Psi_{4,2}+X_{0,2}\Psi_{4,1}=0
\end{eqnarray}
and we recognize $d[J_3]$ to be just the difference of the two first order Dirac currents Eq.(32).
Following the prescription for minimal coupling in Ref. 5, Koshelkin \cite{Kosh1} has set up
and discussed the equations for a threefold SU(2) Yang-Mills
field, which has an EDS generated by forms $\theta^a$ and $\psi_a$, functions of potentials $A^a_i$ and fields $F^a_{ij}$,
when minimally coupled to a multiplet of (two) Dirac fields using
these currents.  The coupling added to the sum of the gauge and bispinor
Cartan 4-forms is $A^a\wedge J_a$.  Taking an exterior derivative, we
find the the multisymplectic 5-form in a 50-dimensional phase space:
\begin{eqnarray}
\text{d$\Lambda $}=-\theta ^1\wedge \left(\psi _1+ J_1\right)-\theta ^2\wedge \left(\psi _2+J_2\right)-\theta^3\wedge \left(\psi _3+ J_3\right) \nonumber \\  +\text{d$\Lambda_{D1} $}+\text{d$\Lambda_{D2} $} +\left(A^1\wedge dJ_1+A^2\wedge dJ_2+A^3\wedge dJ_3\right)
\end{eqnarray}
\begin{equation}
\text{EDS}: \theta^i,  d\theta^i,  ( \psi_i + J_i),  d( \psi_i + J_i), \text{$X \bullet d\Lambda $}
\end{equation}
From this, the EDS should be generated by 2-forms $\theta ^a$, 3-forms $\psi _a+ J_a$ (showing the Dirac bispinor currents as sources of the YM fields) and their closures, together with sixteen 4-forms that are two copies of $\Xi_i$,$\Psi_i$, "covariantly" modified with the YM potential terms in the final parenthesis.  
We calculate the characters to be 50(0,3,9,32,0,0)6, at the same time finding that $x,y,z,t$
are \emph{not} in involution.  Although obtained from a Cartan form, an apparently acceptable variational principle, the EDS is not well-posed.
Koshelkin \cite{Kosh1} \cite{Kosh2}.found a class of solutions to this theory, approximations to
which then lead on to a paradox, so he also concluded that there are in
fact no general self-consistent solutions.  The failure of the well-posedness, and perhaps the need for extra conditions, may not be clear in a perturbative treatment of this coupled SU(2)-bispinor field theory, but the resolution in the case of the standard model was already found by Fadeev and Slavnov. 

In the present case three additional "ghost" fields must be included, also minimally coupled to the gauge currents but without themselves directly contributing to the multisymplectic dynamics.  Using the notation of Section II, instead of Equation (41) and (42) we take

\begin{eqnarray}
\text{d$\Lambda $}=-\theta ^1\wedge \left(\psi _1+ J_1+J_{1s}\right)-\theta ^2\wedge \left(\psi _2+J_2+J_{2s}\right)-\theta^3\wedge \left(\psi _3+ J_3+J_{3s}\right) \nonumber \\  +\text{d$\Lambda_{D1} $}+\text{d$\Lambda_{D2} $} +A^1\wedge (dJ_1+dJ_{1s})+A^2\wedge (dJ_2+dJ_{2s})+A^3\wedge (dJ_3+dJ_{3s})
\end{eqnarray}
\begin{equation}
\text{EDS}: \theta^i,  d\theta^i,  ( \psi_i + J_i+J_{is}),  d( \psi_i + J_i+J_{is}), \text{$X \bullet d\Lambda $}
\end{equation}

The AVF program now finds the Cartan table in 65 dimensions to be 65(0,6,12,37)4+6, well posed, with x, y, z and t in involution and with 6 degrees of gauge freedom, as shown.

\subsection{Summary}
Cartan's formulation of sets of first order partial differential equations as Exterior Differential Systems allows determination of integrability and well-posedness while not resorting to approximation or perturbation theory.  We have used it to explore and characterize some classical coupled theories that are precursers to the standard model of quantum field theory.  QED, or Maxwell-Dirac theory,  proved to be well-posed when written as an EDS induced from a Cartan form set on a state space of 18+4 dimensions. Coupling a SU(2) Yang-Mills field to a Dirac multiplet required three additional scalar "ghost" fields, for a total of 61+4 dimensions, to achieve well-posedness.

(added note:  We have subsequent to  arXiv submission in 2014 found the SU(2) coupled gauge theory of Sec. V as the third auxilliary exercise posed in the online Lecture Notes on Advanced Quantum Field Theory of Claudio Scrucca, Institute of Theoretical Physics, IFPL, Lausanne)

\subsection{Acknowledgements} 
This research was performed while the author held a visiting
appointment at the Jet Propulsion Laboratory, California Institute of Technology.


\begin{thebibliography}{99}
\bibitem{esta}F. B. Estabrook, Exterior Differential Systems for
  Yang-Mills Theories, SIGMA \textbf{4} 063-7 (2008)
\bibitem{Ivey} Ivey T A and Landsberg J M, {\it Cartan for Beginners: Differential Geometry via Moving Frames and Exterior Differential Systems}, (Providence RI: AMS Graduate Studies in Math {\bf 61}, 2003)
\bibitem{Bry} R. Bryant, P. Griffiths and D. Grossman {\it Exterior Differential Systems and Euler-Lagrange Partial Differential Equations} (The University of Chicago Press, Chicago, London, 2003)
\bibitem{Gim} M. J. Gotay, J. Isenberg, J. E. Marsden, and R. Montgomery, Momentum maps and classical relativistic fields. Part I: Covariant field theory, arXiv:physics/9801019.
\bibitem{Got}Gotay M J 1991 An exterior differential system approach to the Cartan
form, Progress in
Mathematics, 99, ed P Donato,  J Elhadad and G M Tuynmen (Boston MA:
Birkh\"{a}user)
\bibitem{cendra}H. Cendra and S. Capriotti, {\it Cartan Algorithm and Dirac Constraints for Griffiths Variational Problems}, arXiv:1309.40801 [math-ph] 16 Sept. 2013

\bibitem{grav}F. B. Estabrook, R. S. Robinson, and H. D. Wahlquist, Constraint-free Theories of Gravitation, \textit{Classical and Quantum Gravity}, \textbf{16}, 911-918 (1999);  F. B. Estabrook, The Hilbert Lagrangian and isometric embedding:  Tetrad formulation of Regge-Teitelboim gravity,  \textit{J. Math. Physics} \textbf{51}, 042502 (2010), arXiv:0908.0365v2 [gr-qc] 
\bibitem{Paston}S. A. Paston and A. A. Sheykin, From the Embedding Theory to General Relativity as a Result of Inflation, \textit{Int. J. Mod. Phys. D}, \textbf{21}, 024024 (2012) arXiv:1105.5212 
\bibitem{xAct}J. M Mart\'{\i}n-Garcia, A. Garcia-Parrado, A.
  Stecchina, B. Wardell, C. Pitrou, D. Brizuela, D. Yllanes, G. Faye,
  L. Stein, R. Portugal, and T. Backdahl, ÒxAct: Efficient tensor
  computer algebra for Mathematica,Ó (GPL 2002-Ð2013),
  http://www.xact.es/.
\bibitem{Bar}A. G. Barut, D. J. Moore and C. Piron, The Cartan
  Formalism in Field Theory, \textit{Helv. Phys. Acta} \textbf{66}
  795-809 (1993).
\bibitem{Fad} L. D. Fadeev and A. A. Slavnov \textit{Gauge Fields:
    Introduction to Quantum Theory} (Ottawa: Benjamin/Cummings, 1980).
\bibitem{Kosh1}A. V. Koshelkin, The Dirac and Gauge Yang-Mills Fields
  in Self-Consistent Configuration, arXiv:1012.0 Dec 2010
\bibitem{Kosh2}A. V. Koshelkin, Solution of Dirac Equation in External
  Yang-Mills Gauge Field, \textit{Physics Lett.} \textbf{683} 205
  (2010)

\end{thebibliography}
\end{document}